\documentclass[aps,preprint,preprintnumbers,nofootinbib,showpacs,superscriptaddress]{revtex4-1}

\usepackage{graphicx,color,amsmath}
\begin{document}
\title{Singularity Problem in Teleparallel Dark Energy Models}
\author{Chao-Qiang Geng}
\email{geng@phys.nthu.edu.tw}
\affiliation{College of Mathematics \& Physics, Chongqing University of Posts \& Telecommunications, Chongqing, 400065, China}
\affiliation{Department of Physics, National Tsing Hua University, Hsinchu, Taiwan 300}
\affiliation{National Center for Theoretical Sciences, Hsinchu, Taiwan 300}
\author{Je-An Gu}
\email{jagu@ntu.edu.tw}
\affiliation{Leung Center for Cosmology and Particle
Astrophysics, National Taiwan University, Taipei, 10617, Taiwan}
\author{Chung-Chi Lee}
\email{g9522545@oz.nthu.edu.tw}
\affiliation{Department of Physics, National Tsing Hua University, Hsinchu, Taiwan 300}
\begin{abstract}
We study future singularity in teleparallel dark energy models,
particularly its behavior and its (non)occurrence in the
observationally viable models.
For the models with a general self-potential of the scalar field, we point out that
both at early times and in the future near the singularity the
behavior of dark energy can be described by the analytic
solutions of the scalar field we obtained for the model with no
self-potential. As to the (non)occurrence in the viable models,
we consider a natural binding-type self-potential, the
quadratic potential, when fitting observational data,
and illustrate the constraining region up to the $3\sigma$
confidence level as well as the region where a singularity will
occur. As a result, the singularity region is outside the
$3\sigma$ constraint. Thus, although the future singularity
problem potentially exists in teleparallel dark energy models,
the observationally viable models may not suffer this problem.
\end{abstract}
\date{\today}
\maketitle

\section{Introduction} \label{sec:introduction}

The late-time acceleration of the cosmic expansion has been confirmed by
a variety of observational data, such as those from type-Ia supernovae (SNIa)~\cite{astro-ph/9805201, astro-ph/9812133}, cosmic microwave background radiation (CMB)~\cite{astro-ph/0302209, Komatsu:2010fb, Ade:2013xsa} and baryon acoustic oscillation (BAO)~\cite{Eisenstein:2005su}.
This salient phenomenon may be explained simply by a cosmological constant, or suggest the existence of a new dynamical degree of freedom, either in the energy contents or in gravity, that provides anti-gravity. The origin and the nature of such anti-gravity is one of the most important problems in cosmology and astrophysics.

The new degree of freedom in the energy contents is generally called ``dark energy''~\cite{Copeland:2006wr}, while that in gravity gives a modification of gravity. The simplest degree of freedom is a scalar field, which is called quintessence when minimally coupled to gravity.
In the present paper we will consider teleparallel dark energy, a  dynamical scalar field non-minimally coupled to teleparallel gravity. It can be regarded as a new degree of freedom both in the energy contents and in the modification of gravity.
The teleparallel dark energy model as an extension of teleparallel gravity is analogous to the minimal extension of the quintessence model in general relativity, i.e.\ the scalar-tensor theory, but has a richer
structure~\cite{Geng:2011aj, Geng:2011ka, Wei:2011yr, Xu:2012jf, Gu:2012ww, Banijamali:2012nx, Aslam:2012tj, Geng:2012vn, Wang:2013wuf, Sadjadi:2013nb, Otalora:2013dsa}.

We have studied in Ref.~\cite{Gu:2012ww} the teleparallel dark energy model with no potential but simply the non-minimal coupling. We derived the analytic solutions of the scalar field in the radiation dominated (RD), matter dominated (MD) and scalar field/dark energy dominated (SD) eras, respectively. In SD we found a finite-time singularity that the Hubble expansion rate will go to infinity at a finite scale factor $a_s$ and a finite value of the scalar field $\phi_s$.
%This is an unexpected result in the cosmological model.
This is caused by the non-minimal coupling that effectively changes the gravitational coupling strength and can even make it diverge
when $\phi$ is driven to some specific value $\phi_s$.
% tends to drive the scalar field to infinity and makes the scalar field
% reach $\phi_s$ where the singularity occurs.

In the present paper we study the future singularity problem in
teleparallel dark energy models with a self-potential of the
scalar field. In particular, we investigate (1) the behavior of
the future singularity in the models with a general
self-potential and (2) the (non)occurrence of the future singularity
in the observationally viable models with a binding-type
potential that provides an opportunity to avoid the singularity
by confining the scalar field and keeping it away from
$\phi_s$. We will fit observational data, obtain the
observational constraints of the model, and then examine
whether the constraining region overlaps with the singularity
region. If no overlap, the singularity problem does not appear
in the data-favored region of the model, although it
potentially exists in teleparallel dark energy models.
%%%%%%
% For the former we point out that the behavior of dark energy
% near the singularity can be described by the analytic solution
% we obtained in Ref.~\cite{Gu:2012ww} for the model with no
% self-potential.
% For the latter we consider the model with a quadratic
% potential, a natural binding-type potential, and perform the
% data fitting. We will illustrate both the constraining region
% up to the $3\sigma$ confidence level and the region where a
% singularity will occur, thereby showing whether the
% observationally constraining region of the model can be free of
% singularity.
% If these two regions do not overlap, the problem of future
% singularity does not appear in the data-favored region of the
% model space, although it potentially exists in teleparallel
% dark energy models.
%%%%%%

This parer is organized as follows. In Sec.~\ref{sec:TeleDE} we
introduce the teleparallel dark energy model with a general
potential and study the singularity problem analytically. In
Sec.~\ref{sec:sing} % we first classify the type of the singularity
we numerically analyze a quadratic potential (binding-type) and
an exponential potential (unbinding-type), and fit the former
to observational data. A summary is given in
Sec.~\ref{sec:summary}.

\section{Teleparallel Dark Energy Model} \label{sec:TeleDE}

In the theory of teleparallel gravity, gravity is described by torsion instead of curvature, and the dynamical variable is the vierbein field $\mathbf{e}_a (x^{\mu})$ (also called tetrad) that forms an orthonormal coordinate at each space-time point $x^{\mu}$,
\begin{eqnarray}
\mathbf{e}_a \cdot \mathbf{e}_b = \eta_{ab} =  \mathrm{diag}(1,-1,-1,-1) \,.
\end{eqnarray}
The metric is given by the vierbein field as %
$g_{\mu \nu}= \eta_{a b} e^a_{\ \mu} e^b_{\ \nu}$.
%\begin{eqnarray}
%g_{\mu \nu}= \eta_{a b} e^a_{\ \mu} e^b_{\ \nu} \,.
%\end{eqnarray}
The torsion tensor is defined as the anti-symmetric part of the connection,
% which exchange their lower indices:
\begin{eqnarray}
\label{eq:torsion_tensor}
T^{\lambda}_{\ \mu \nu}= \overset{\mathbf{w}}{\Gamma}{}^{\lambda}_{\ \nu \mu} - \overset{\mathbf{w}}{\Gamma}{}^{\lambda}_{\ \mu \nu} = e^{\ \lambda}_a \partial_{\mu} e^a_{\ \nu} - e^{\ \lambda}_a \partial_{\nu} e^a_{\ \mu} \, ,
\end{eqnarray}
where the Weitzenb\"ock connection $\overset{\mathbf{w}}{\Gamma}{}^{\lambda}_{\ \nu \mu} \equiv e^{\ \lambda}_a \partial_{\mu} e^a_{\ \nu}$. The gravity Lagrangian with teleparallelism is given by the torsion scalar,
\begin{eqnarray}
\label{eq:torsion_scalar}
T = \frac{1}{4}T^{\rho \mu \nu }T_{\rho \mu \nu }
+ \frac{1}{2}T^{\rho \mu \nu}T_{\nu \mu \rho }
- T_{\rho \mu }^{\ \ \rho }T_{\ \ \ \nu }^{\nu \mu} \, .
\end{eqnarray}

The teleparallel dark energy model invokes a scalar field which is non-minimally coupled to teleparallel gravity. Its action reads
\begin{equation}
\label{eq:action_TDE}
S=\int d^{4}x \, e \left[ \frac{T}{2\kappa^{2}}
+ \frac{1}{2} \left(\partial_{\mu}\phi\partial^{\mu}\phi+\xi
T\phi^{2}\right) - V(\phi) +\mathcal{L}_m \right] ,
\end{equation}
where $e\equiv \text{det}(e_{\ \mu}^a) = \sqrt{-g}$, $\kappa$ and $\xi$ are coupling constants, $V(\phi)$ is the self-potential of the scalar field and $\mathcal{L}_m$ the matter Lagrangian.
%%%%%% %%%%%% %%%%%%
%Varying the action with respect to the vierbein field gives the gravitational field equation~\cite{Geng:2011aj}:
%\begin{eqnarray}\label{eq:eom1}
%&&\left(\frac{2}{\kappa^2}+2 \xi
%\phi^2 \right)\left[e^{-1}\partial_{\mu}(ee_A^{\rho}S_{\rho}{}^{\mu\nu} )
%-e_{A}^{\lambda}T^{\rho}{}_{\mu\lambda}S_{\rho}{}^{\nu\mu}
%-\frac{1}{4}e_{A}^{\nu}T\right]\nonumber\\
%&&- e_{A}^{\nu}\left[\frac{1}{2}
%\partial_\mu\phi\partial^\mu\phi-V(\phi)\right]+
%  e_A^\mu \partial^\nu\phi\partial_\mu\phi
%+ 4\xi e_A^{\rho}S_{\rho}{}^{\mu\nu}\phi
%\left(\partial_\mu\phi\right)
%=e_{A}^{\rho}\overset {\mathbf{em}}T_{\rho}{}^{\nu},
%\end{eqnarray}
%where $\overset{\mathbf{em}}{T}_{\rho}{}^{\nu}$ stands for the usual
%energy-momentum tensor.
%%%%%% %%%%%% %%%%%%
For a flat, homogeneous and isotropic universe where $\phi = \phi(t)$ and $g_{\mu \nu} =\mathrm{diag}(1, -a^2, -a^2, -a^2)$, the vierbein field $e^b_{\ \mu} =\mathrm{diag}(1, a, a, a)$, and the scalar and gravitational field equations derived from the above action read~\cite{Geng:2011aj}
\begin{eqnarray}
\label{eq:eom2}
&& \ddot{\phi} + 3H\dot{\phi} + 6\xi H^2\phi +V_{,\phi} = 0 \, ,  \\
&& H^{2}=\frac{\kappa^2}{3} \left( \rho_{\phi}+\rho_{m}+\rho_{r} \right),
\label{eq:eom_fried1} \\
&& \dot{H}=-\frac{\kappa^2}{2} \left( \rho_{\phi}+p_{\phi}+\rho_{m}+4\rho_{r}/3
\right).
\label{eq:eom_fried2}
\end{eqnarray}
Here the Hubble expansion rate $H\equiv\dot{a}/a$, the energy
density of non-relativistic matter $\rho_m \propto a^{-3}$,
that of radiation $\rho_r \propto a^{-4}$, and the energy
density and pressure of the scalar field
\begin{eqnarray}
\label{eq:rho_phi}
\rho_{\phi} &=& \frac{1}{2} \dot{\phi}^2 + V(\phi) - 3\xi H^2 \phi^2 \, , \\
\label{eq:p_phi}
p_{\phi} &=& \frac{1}{2} \dot{\phi}^2 - V(\phi) + 3\xi H^2 \phi^2 + 2\xi \frac{d}{dt}(H\phi^2) \, .
\end{eqnarray}
The equation of state (EoS) of the scalar field is defined as
$w_{\phi} = p_{\phi} / \rho_{\phi}$. %
The gravitational field equations can be rewritten as
\begin{eqnarray}
H^2 &=& \frac{1}{3} \left(\frac{\kappa^2}{1+\xi\kappa^2\phi^2}\right)
\left( \frac{1}{2}\dot{\phi}^2 + V + \rho_m + \rho_r \right),
\label{eq:Fried1Rewritten} \\
-\dot{H} &=& \frac{1}{2} \left(\frac{\kappa^2}{1+\xi\kappa^2\phi^2}\right)
\left( \dot{\phi}^2 + 4\xi H\phi\dot{\phi}
+ \rho_m + 4\rho_r/3 \right),
\label{eq:Fried2Rewritten}
\end{eqnarray}
where $\kappa^2/(1+\xi\kappa^2\phi^2)$ may be regarded as the effective gravitational coupling strength.

For the singularity problem we will consider the case with
negative $\xi$ as in Ref.~\cite{Gu:2012ww}. In this case the
non-minimal coupling term in the scalar field equation
(\ref{eq:eom2}) tends to drive the scalar field to infinity.
The above equations show that
a singularity occurs when $\kappa\phi$ is driven to
$\kappa\phi_s \equiv \pm 1/\sqrt{-\xi}$ where the effective
gravitational coupling strength goes to infinity.
In Eqs.\ (\ref{eq:eom_fried1})--(\ref{eq:p_phi}) both a
positive potential and the term $3 \xi H^2 \phi^2$ contribute
negative pressure (while the term $2\xi d(H\phi^2)/dt$ is
undetermined).
Accordingly, the non-minimal coupling % between the scalar field and teleparallel gravity
with a negative coupling constant $\xi$ may provide repulsive
gravitation onto the universe as well as a ``repulsive force''
onto the scalar field.

\section{Singularity in Teleparallel Dark Energy Models} \label{sec:sing}

In this section we investigate the behavior of the future singularity in the
teleparallel dark energy model. According to the behavior of the scale factor, the effective dark energy density $\rho_\mathrm{eff}$ and pressure $p_\mathrm{eff}$ at the singularity, the future singularities have been classified \cite{Nojiri:2005sx} as follows.
When $t\rightarrow t_{s}$,
\begin{itemize}
\item \noindent \textbf{ Type I (Big Rip):}
    $a(t)\rightarrow\infty$,
    $\rho_\mathrm{{eff}}\rightarrow\infty$ and
    $|p_\mathrm{{eff}}|\rightarrow\infty$,
\item \noindent \textbf{ Type II (Sudden):}
 $a(t)\rightarrow a_{s}$,
$\rho_\mathrm{{eff}}\rightarrow\rho_{s}$ and
$|p_\mathrm{{eff}}| \rightarrow\infty$,
\item \noindent \textbf{ Type III:} $a(t)\rightarrow
    a_{s}$, $\rho_\mathrm{{eff}}\rightarrow\infty$ and
    $|p_\mathrm{{eff}}|\rightarrow\infty$,
\item \noindent \textbf{ Type IV:} $a(t)\rightarrow a_{s}$,
$\rho_\mathrm{{eff}}\rightarrow 0$, $|p_\mathrm{{eff}}|
\rightarrow 0$ and higher derivatives of $H$ diverge,
\end{itemize}
where $t_s$ and $a_s$ are finite.
% and $\rho_\mathrm{eff}$ and $p_\mathrm{eff}$
% are respectively the effective dark energy density and pressure.
Recent studies on future singularity for other models related to the torsion scalar can be
found in Refs.~\cite{Bamba:2012ka,Bamba:2012vg}.

In the following we will first analytically study the models
with a general potential, then numerically analyze two specific
types of potentials, and perform data fitting for the
binding-type potential that has the opportunity to avoid the
singularity.

\subsection{Teleparallel dark energy with a general potential}
In the viable models consistent with data, the potential is
generally negligible at early times. Accordingly, in RD and MD
the scalar field is mainly driven by the non-minimal coupling,
% i.e., $6\xi H^2 \phi$ in Eq.~(\ref{eq:eom2}),
and can be approximately described by the analytic solutions
we obtained in Ref.~\cite{Gu:2012ww} for the model with no
potential. The solutions are presented as follows.

In RD and MD the Hubble expansion rate $H = \alpha/t$, where
the constant $\alpha = (2/3)(1+w_D)^{-1}$ and $w_D$ is the
constant EoS of the dominant energy source.
% $w_{eff}=1/3$ in RD and $w_{eff}=0$ in MD
When $V(\phi)$ is negligible, the solution of the scalar field
in these two eras is a linear combination of two modes:
\begin{equation}
\phi (t) = C_1 t^{l_1} + C_2 t^{l_2} \, ,
\end{equation}
where $C_{1,2}$ are integration constants and
\begin{equation}
l_{1,2} =  % \hspace{0.7em}
\frac{1}{2} \left[ \pm \sqrt{(3\alpha-1)^2 - 24\xi \alpha^2} - (3\alpha-1) \right].
\label{eq:sol_phi1}
\end{equation}
With negative $\xi$, $l_1$ and $l_2$ are positive and negative,
respectively. Therefore, the $l_1$ mode, as an increasing mode,
will soon dominates over the other decreasing ($l_2$) mode.

Henceforth we simply use the $l_1$ mode to describe the
scalar field. The EoS of this mode
\begin{equation}
\label{eq:EoS_phi}
w_{\phi} % \equiv \frac{p_{\phi}}{\rho_{\phi}}
= -1+\frac{2(1-l_1)}{3\alpha} \, ,
% = -1 + (1+w_{eff})(1-l_1)
% < w_{D} = -1 + \frac{2}{3\alpha} \, ,
\end{equation}
which is independent of the initial condition, a tracker behavior
pointed out in Ref.~\cite{Gu:2012ww}. We note that $w_{\phi}$
is always smaller than that of the dominant energy
source, $w_{D} = -1 + 2/(3\alpha)$, if $w_D > -1$. This
guarantees the existence of the late-time SD following MD,
a generic feature of the teleparallel dark energy model.

At the late times when dark energy becomes important, the
self-potential $V(\phi)$ competes with the
non-minimal coupling to teleparallel gravity for the evolution
of the scalar field. If $V(\phi)$ can prevent $\phi$ from
reaching the singularity point $\phi_s$, the singularity can be avoided.
On the contrary, if $\phi$ is still driven to $\phi_s$ even under $V(\phi)$,
the singularity will occur.

Around the singularity, because of the extremely rapid expansion,
$V(\phi)$, $\rho_m$ and $\rho_r$ can be ignored  in the Friedmann
equation when compared with the term proportional to
$H^2\phi^2$. Equation (\ref{eq:Fried1Rewritten}) then gives
\begin{equation}
-(\kappa\phi^{\prime})^2/6 + \xi(\kappa\phi)^2 + 1 \approx 0 \, ,
\end{equation}
where the prime denotes the derivative with respect to the
number of e-folding, $N\equiv \ln a$. In this case the behavior
of the scalar field can be approximately described by the
analytic solution $\phi(N)$ in SD we obtained in Ref.~\cite{Gu:2012ww}
with no potential:
\begin{eqnarray}
\phi(N) &=& \pm \sin\theta(N)/\sqrt{-\xi} \, , \\
\theta(N) &\equiv& \sqrt{-6\xi}N + C \, ,
\end{eqnarray}
where $C$ is an integration constant and $\theta$ linearly
increases with the e-folding number of the cosmic expansion.
The dark energy EoS
\begin{equation}
w_\phi = -1 - \sqrt{-32\xi/3} \tan\theta \, .
\end{equation}
As a result, when the cosmic scale factor increases to the value $a_s$ at which
$\sin\theta = \pm 1$, $\kappa\phi = \kappa\phi_s \equiv
\pm 1/\sqrt{-\xi}$ and $\phi^{\prime} = 0$, the universe
meets a type-III singularity:
\begin{equation}
H,\rho_\phi \rightarrow \infty \, , \quad
p_\phi , w_\phi \rightarrow -\infty \, .
\end{equation}

\subsection{Numerical analysis}

To demonstrate the features of teleparallel dark energy
pointed out in the above analytical study,
in this section we consider two specific potentials, a
quadratic and an exponential potential, that respectively
represent the binding-type and the unbinding-type
potentials. We will show how the former may avoid the
singularity while the latter cannot. We will then fit the
former to observational data and thereby examine whether
the singularity can be avoided in the observationally viable
region of this model.

We write the quadratic potential as $V(\phi) = p V_0
(\kappa\phi)^2$, where $V_0 = \rho_m^{(0)}/3$, i.e.\ one third
of the present matter energy density, and $p$ is a
dimensionless constant and the only free parameter of the
potential. This self-interaction potential tends to confine the
scalar field around zero, while the non-minimal coupling tends
to drive the scalar field to infinity. The competition between
them determines whether $\phi$ can reach the singularity point
$\phi_s$, i.e., whether the singularity will occur.

\begin{figure}[h!]
\includegraphics[width=6cm]{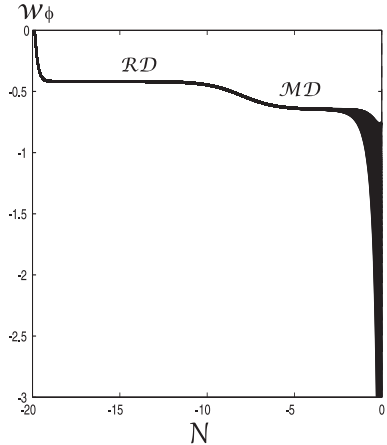}
\caption{\label{fg:loga-w01}
The evolution of the dark energy EoS for $V(\phi)=pV_0 (\kappa\phi)^2$
with $(p,\xi)=(2,-0.4)$ and with the initial conditions: $\kappa \phi_{(i)} \in [10^{-14},2 \times 10^{-9}]$ and $\phi^{\prime}_{(i)}=0$ at $N=-20$. The black area is formed by the curves with respect to this wide range of initial conditions.}
\end{figure}
\begin{figure}[h!]
\includegraphics[width=6cm]{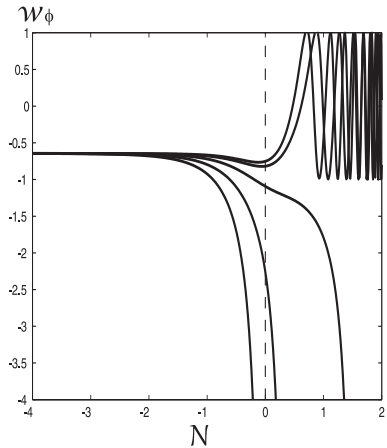}
\caption{\label{fg:loga-w02}
Five initial conditions are picked to present the more detailed evolution of the dark energy EoS in the same quadratic-potential model as Fig.~\ref{fg:loga-w01}: $ \kappa \phi_{(i)} = 10^{-14}$, $5\times 10^{-10}$, $10^{-9}$, $1.5 \times 10^{-9}$ and $2 \times 10^{-9}$ (from top to bottom) at $N=-20$. The dashed line denotes the present time.}
\end{figure}
\begin{figure}[h!]
\includegraphics[width=6cm]{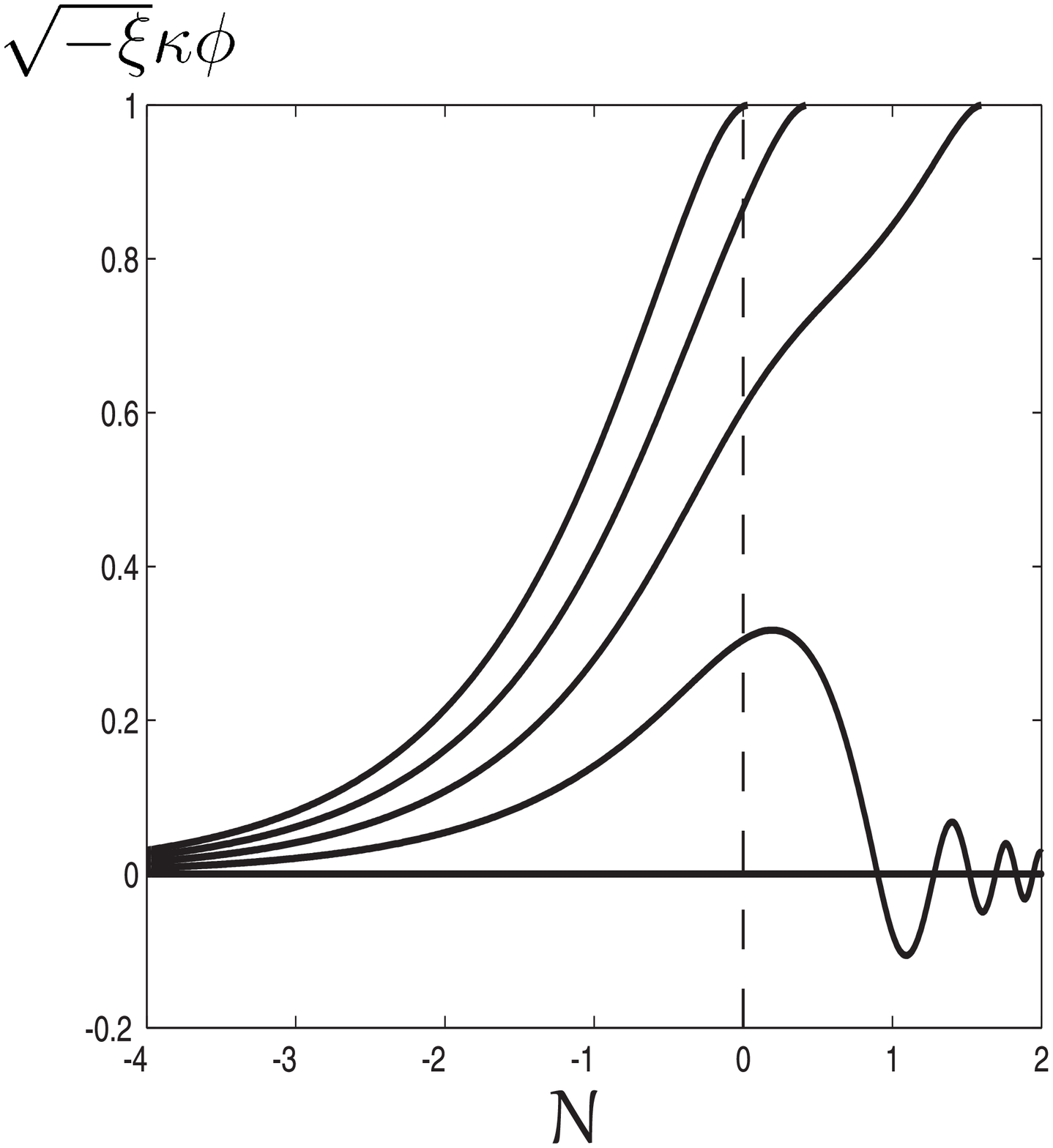}
\includegraphics[width=6cm]{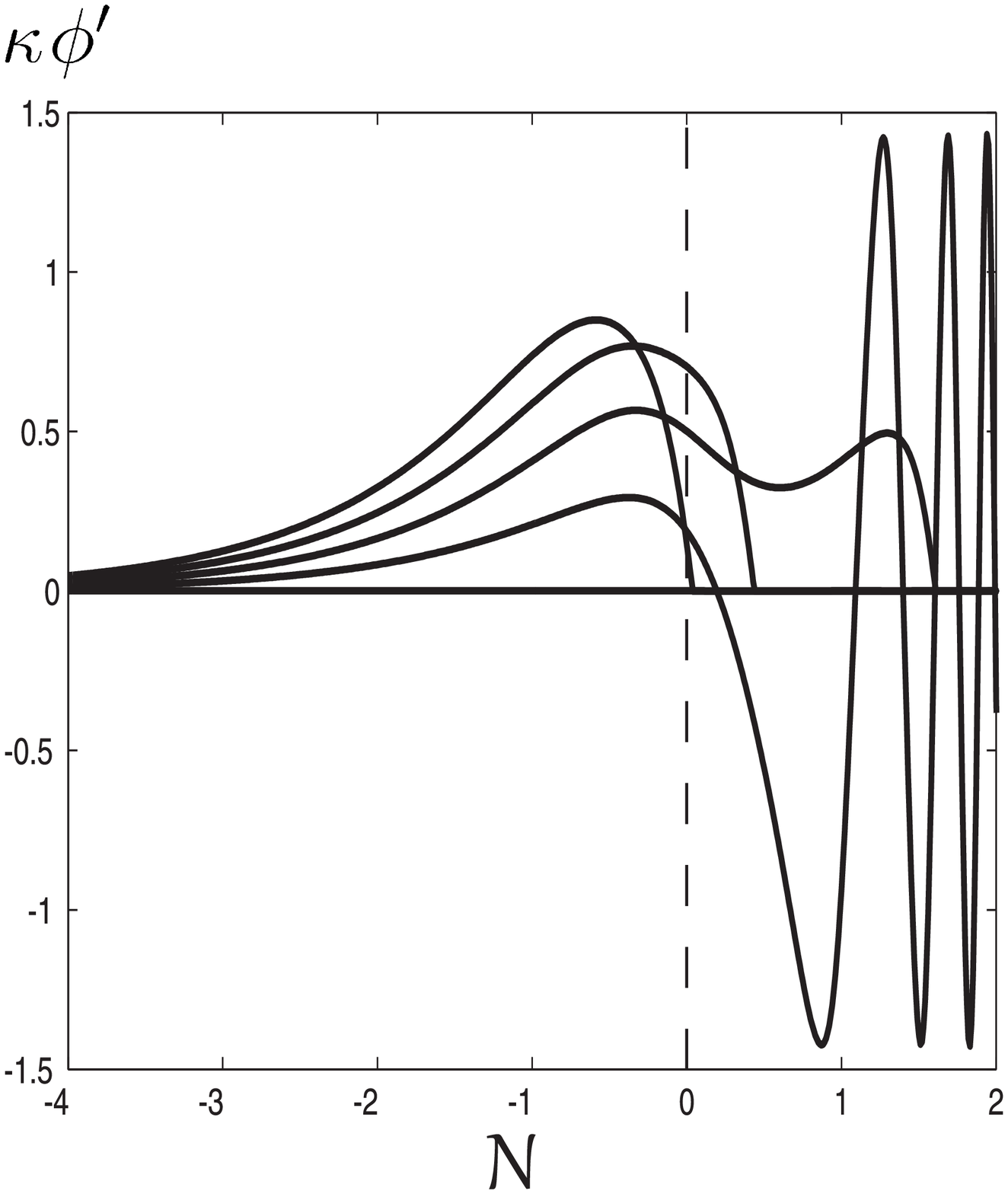}
\caption{\label{fg:loga-psi01}
The evolution of $\phi$ and $\phi^{\prime}$ in the same quadratic-potential model  as Fig.~\ref{fg:loga-w01} with the same five initial conditions picked in Fig.~\ref{fg:loga-w02}. (The initial value $\phi_{(i)}$ decreases from top to bottom.)}
% The dashed line denotes the present time.
\end{figure}

As an example for demonstration, here we set $p=2$ and
$\xi=-0.4$, and numerically obtain the evolution of the scalar
field
% along with the number of e-folding of the cosmic expansion $N$
for a wide range of initial conditions: $\kappa \phi_{(i)} \in
[10^{-14},2 \times 10^{-9}]$ and $\phi_{(i)}^{\prime}=0$ at
$N=-20$.
%%%%%% 2013.07.02 revised %%%%%%%%%%%%%%%%%%%%%%%%%%%%%%%%%%%%%%
This setting is made according to the phenomenological
requirement that $\Omega_{\phi 0} \sim \mathcal{O}(1)$ and
$w_{\phi 0} \sim \mathcal{O}(-1)$.
%%%%%%%%%%%%%%%%%%%%%%%%%%%%%%%%%%%%%%%%%%%%%%%%%%%%%%%%%%%%%%%%
The evolution of the dark energy EoS along with the e-folding
number of the cosmic expansion, $w_{\phi}(N)$, is presented in
Fig.~\ref{fg:loga-w01}. Among these initial conditions, we pick
five of them and present the more detailed evolution of
$w_{\phi}$ in Fig.~\ref{fg:loga-w02} and the evolution of
$\phi$ and $\phi^{\prime}$ in Fig.~\ref{fg:loga-psi01}.
%%%%%% 2013.07.02 revised %%%%%%%%%%%%%%%%%%%%%%%%%%%%%%%%%%%%%%
The occurrence of a future singularity is indicated when
$w_{\phi}$ goes down rapidly, e.g., the lower three curves in
Fig.~\ref{fg:loga-w02}.
%%%%%%%%%%%%%%%%%%%%%%%%%%%%%%%%%%%%%%%%%%%%%%%%%%%%%%%%%%%%%%%%

These three figures manifest the features we have pointed out
in the above analytical study: the tracker behavior of
$w_{\phi}$, the late-time dominance of dark energy,
and the possibility of avoiding the singularity.
This possibility depends on the initial condition:
For a larger initial value $\phi_{(i)}$,
i.e., closer to the singularity point $\phi_s$,
the universe enters SD earlier and
will meet the singularity even under
a binding-type potential, as shown by the lower three curves in
Fig.~\ref{fg:loga-w02} and the upper three in
Fig.~\ref{fg:loga-psi01}. The singularity occurs when
$\sqrt{-\xi} \kappa \phi = 1 $ and $\phi^{\prime}=0$ at $\ln
a_s \simeq 0.04$, $0.44$ and $1.61$, respectively. %
In contrast, for a smaller initial value $\phi_{(i)}$ the
scalar field will be pulled back by the potential before
reaching $\phi_s$; thereafter, both the scalar field and its
EoS oscillate around zero, i.e.\ behaving as a massive field,
as shown by the upper two curves in
Fig.~\ref{fg:loga-w02} and the lower two in
Fig.~\ref{fg:loga-psi01}.\footnote{The horizontal line in
Fig.~\ref{fg:loga-psi01} shows the evolution for the smallest
initial value $\phi_{(i)}$. It oscillates at late times, but
the amplitude is too small to be seen.}

\begin{figure}[h!]
\includegraphics[width=6cm]{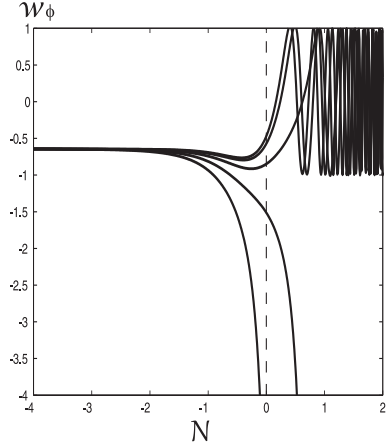}
\caption{\label{fg:loga-w03}
The evolution of the dark energy EoS in the quadratic-potential model with $p=5$ and with the same five initial conditions in Fig.~\ref{fg:loga-w02}.}
\end{figure}
%

%%%%%%
% The threshold from the singularity mode to the massive mode, of
% course, depends on the depth ($p$) of the potential.
%%%%%%
In addition to the initial condition, the value of $p$ is
another key to the singularity. A larger $p$ corresponds to a
steeper potential and therefore a larger ``binding force'' on
$\phi$ that gives a better chance to avoid the singularity.
This feature is illustrated in Fig.~\ref{fg:loga-w03} where we
consider a larger value, $p=5$, and present $w_{\phi}(N)$ for
the same five initial conditions as Fig.~\ref{fg:loga-w02}.
This figure shows (i) the curve in the middle is
previously singular but now becomes non-singular, (ii) the
upper two non-singular curves enter the oscillating phase
earlier, and (iii) the lower two singular curves enter SD
and then meet the singularity later than
the previous case with $p=2$.

\begin{figure}[h!]
\includegraphics[width=6cm]{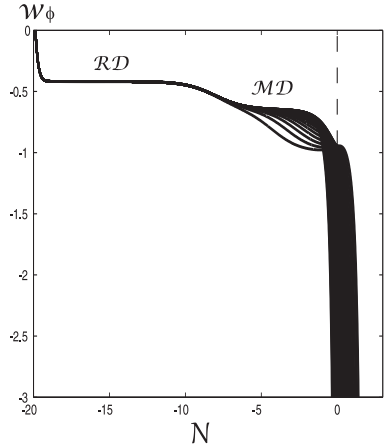}
\caption{\label{fg:exp-x}
The evolution of the dark energy EoS for $V(\phi)=p V_0 e^{-\kappa \phi}$
with $(p,\xi)=(2,-0.4)$ and with the initial conditions: $\kappa \phi_{(i)} \in [10^{-10}, 2 \times 10^{-9}]$ and $\phi^{\prime}_{(i)}=0$ at $N=-20$.}
\end{figure}

For comparison, here we consider an unbinding-type potential,
$V(\phi)=p V_0 e^{-\kappa \phi}$. We choose $p=2$ and the
initial conditions: $\kappa \phi_{(i)} \in [10^{-10}, 2 \times
10^{-9}]$ and $\phi^{\prime}_{(i)}=0$ at $N=-20$. This
unbinding potential apparently does not provide a binding force
against the non-minimal coupling when the scalar field is
driven towards $+\infty$.
%%%%%% 2013.07.02 revised %%%%%%%%%%%%%%%%%%%%%%%%%%%%%%%%%%%%%%
Therefore the singularity will always occur, as shown by
Fig.~\ref{fg:exp-x} where $w_{\phi}$ always goes down rapidly
after MD.
%%%%%%%%%%%%%%%%%%%%%%%%%%%%%%%%%%%%%%%%%%%%%%%%%%%%%%%%%%%%%%%%
%%%%%%
% As a result, there is no matter dominated tracker behavior, but
% the singularity still exists in the future. Hence, to avoid the
% singularity problem one should exclude these un-bounded potentials.
%%%%%%

%
\begin{figure}[h!]
\includegraphics[width=6cm]{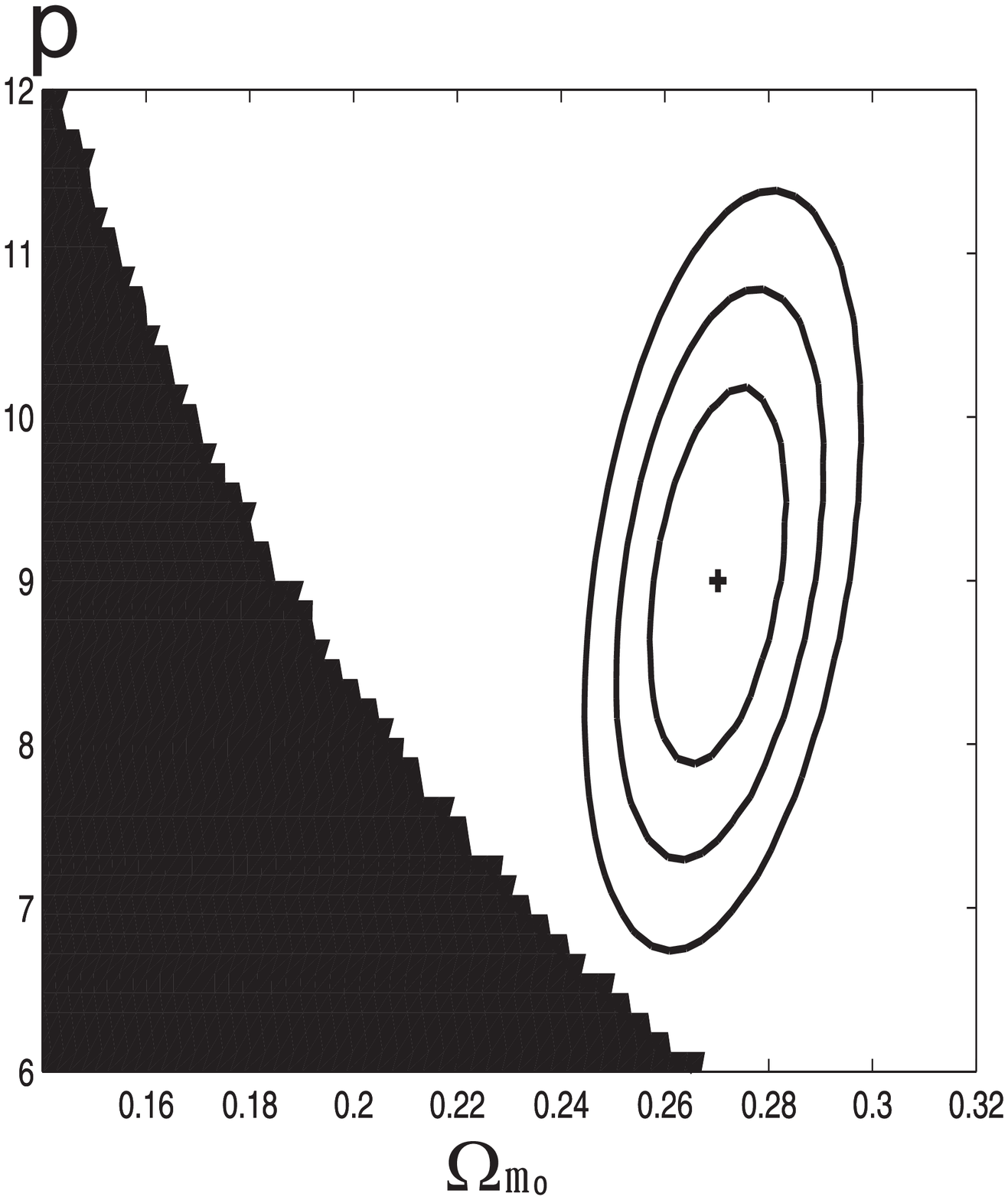}
\includegraphics[width=6cm]{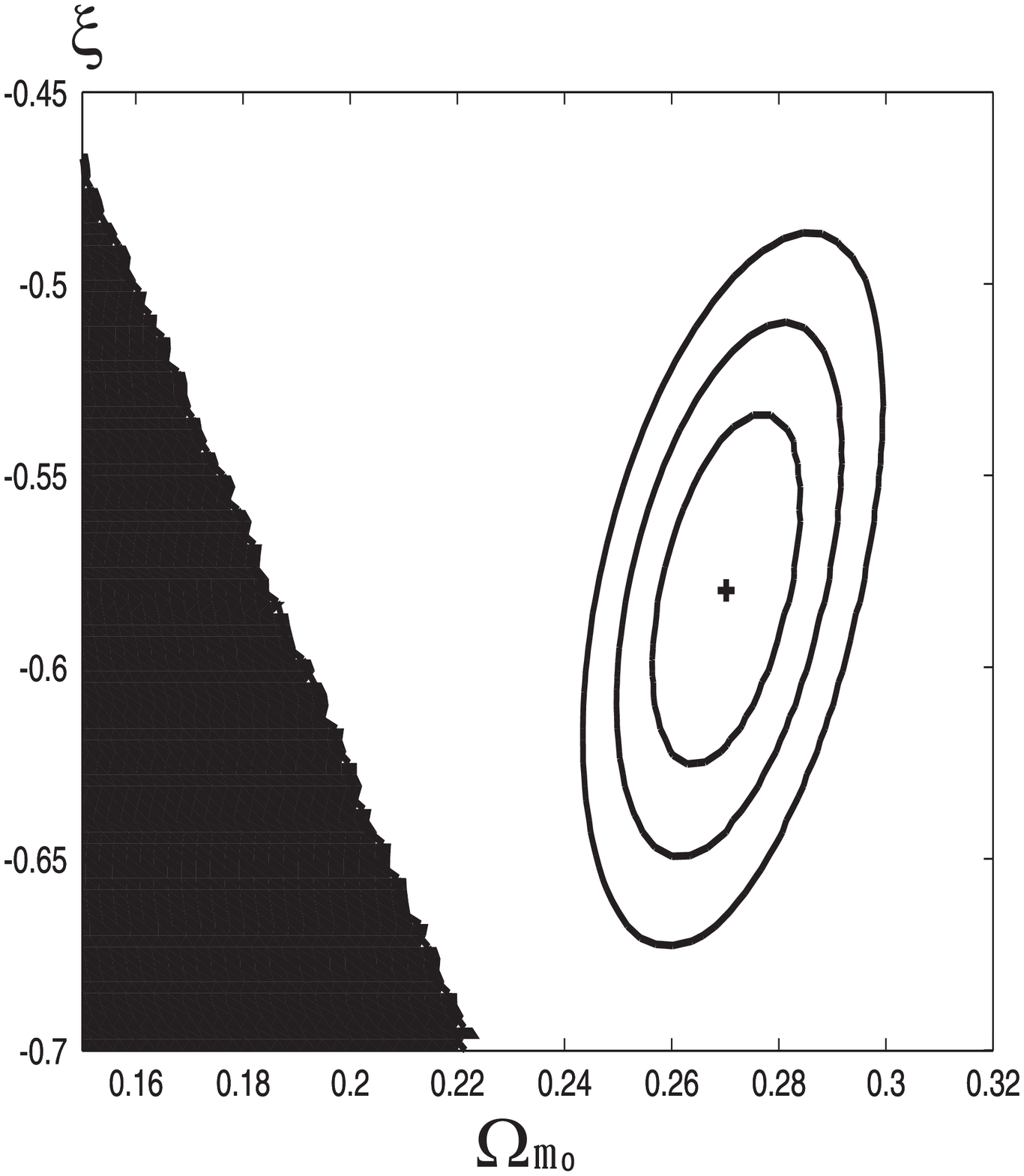}
\caption{\label{fg:contours}
The $1\sigma$--$3\sigma$ constraints on the quadratic-potential model, where $\xi=-0.58$ and $p=9.0$ are respectively set in the left and the right panel. The black area denotes the region where a future singularity will occur.}
\end{figure}

To examine the cosmological viability of the teleparallel dark
energy models with no singularity, we fit the
quadratic-potential model to the SNIa, BAO and CMB data,
following the procedure in Ref.~\cite{Geng:2011ka}. The result
is presented in Fig.~\ref{fg:contours}, where three contours
show the $1\sigma$--$3\sigma$ constraints on the present matter
energy density fraction $\Omega_{m0}$ together with the model
parameters $p$ and $\xi$; the black area denotes the region
where a singularity will occur. This figure shows that the
observationally viable region up to the $3\sigma$ confidence
level in this model is free of singularity. We note that the
$\chi^2$ value of the best-fit of this model is $566.0$ that is
smaller than that of the $\Lambda$CDM model,
$\chi^2_{\Lambda\mathrm{CDM}} = 567.5$.
%%%%%% 2013.07.02 revised %%%%%%%%%%%%%%%%%%%%%%%%%%%%%%%%%%%%%%
In addition, the reduced $\chi^2$ values of these two models
are both around $1.01$, although $\Lambda$CDM invokes fewer
parameters to give a better chance of having a smaller reduced
$\chi^2$ value. This is because the current data favor the
crossing of the phantom divide line ($w_{\phi}=-1$)
from the phantom phase ($w_{\phi}<-1$) to quintessence one ($w_{\phi}>-1$) as the redshift increases,
%regarding the late-time behavior of dark energy, 
a characteristic of the teleparallel dark energy~\cite{Geng:2011aj, Geng:2011ka} as well as other $f(T)$ models~\cite{Bamba2011}.
%%%%%%%%%%%%%%%%%%%%%%%%%%%%%%%%%%%%%%%%%%%%%%%%%%%%%%%%%%%%%%%%
As a result, this model can fit the observational data well and
the model in the fitting region suffers no future singularity
problem.

\section{Summary} \label{sec:summary}

In this paper we study the future singularity in the
teleparallel dark energy models with a self-potential of the
scalar field. A future singularity may appear due to the
non-minimal coupling of the scalar field to teleparallel
gravity that tends to drive the scalar field to infinity. This
singularity may be avoided by a binding-type potential that
tends to confine the scalar field around a finite value. The
destiny is determined by the competition between the
self-interaction and the non-minimal coupling.

For the model with a general potential that looks difficult to
analyze, we point out that the potential may be negligible at
early times in the observationally viable models, as well as in
the future near the singularity (if it exists). Therefore, in
these epochs the scalar field can be approximately described by
the analytic solutions in RD, MD and SD we obtained in
Ref.~\cite{Gu:2012ww} for the model with no potential. These
analytic solutions show the tracker behavior of the dark energy
EoS at early times (RD and MD), the inevitability of the
late-time SD, and the possibility of meeting a type-III
singularity in the future.

To demonstrate the possibility of avoiding the future
singularity under a binding-type potential, we numerically
analyze the model with a quadratic potential. With the
numerical results we illustrate the above features read from
the analytic solutions, and show how the (non)occurrence of the
future singularity depends on the initial conditions and the
steepness of the potential, both of which affect the
competition between the self-interaction and the non-minimal
coupling.

To examine whether the (non)occurrence of the singularity is
favored by observational data, we fit the quadratic-potential
model to the SNIa, CMB and BAO data. We present the
$1\sigma$--$3\sigma$ constraints on the cosmological and the
teleparallel dark energy model parameters, including the
present matter energy density fraction, the non-minimal
coupling constant and the mass scale of the quadratic potential
(corresponding to the steepness of the potential). In addition,
we illustrate in this parameter space the region where a future
singularity will occur. As a result, the singularity region is
outside the $3\sigma$ constraint, i.e., the current data favor
the model region with no singularity. Thus, although the
teleparallel dark energy models potentially have the future
singularity problem, the observationally viable models may
suffer no such problem.

%\vspace{0.5em}
%%%%%%%%%%%%%%%%%%%%%%%%%%%%%%%%%%%%%%%%%%%%%%%%%%%%%%%%%%%%%%%%%%%%%%%
\begin{acknowledgments}
The work was supported in part by National Center for
Theoretical Sciences, National Science Council
(NSC-98-2112-M-007-008-MY3, NSC-101-2112-M-002-007 and
NSC-101-2112-M-007-006-MY3) and 
NTHU
%National Tsing-Hua University
(102N1087E1 and 102N2725E1), Taiwan, R.O.C.
\end{acknowledgments}
%%%%%%%%%%%%%%%%%%%%%%%%%%%%%%%%%%%%%%%%%%%%%%%%%%%%%%%%%%%%%%%%%%%%%%%

%

\end{document}